\documentclass[aps,prl,floatfix,twocolumn,reprint,amsmath,amssymb,superscriptaddress]{revtex4-1}
\usepackage{amsfonts}
\usepackage{mathrsfs}
\usepackage{amsmath}
\usepackage{color}
\usepackage{natbib}
\usepackage{graphicx}
\usepackage{bm}
\usepackage{amssymb}
\usepackage{xspace}
\usepackage{epstopdf}
\usepackage{dcolumn}
\usepackage{longtable}
\usepackage{multirow}
\usepackage[colorlinks=true, letterpaper=true, pdfstartview=FitV, linkcolor=blue, citecolor=blue, urlcolor=blue]{hyperref}
\UseRawInputEncoding

\makeatletter

\newcommand{\Rmnum}[1]{\expandafter\@slowromancap\romannumeral #1@}
\makeatother

\begin{document}

\title{Type-III Weyl Semi-Half-Metal in an Ultralight Monolayer Li$_2$N}

\author{Qingqing Li}
\affiliation{College of Physics and Electronic Engineering, Center for Computational Sciences, Sichuan Normal University, Chengdu, 610068,
P.R.China}

\author{Li Chen}
\affiliation{College of Physics and Electronic Engineering, Center for Computational Sciences, Sichuan Normal University, Chengdu, 610068,
P.R.China}

\author{Run-Wu Zhang}
\email[]{zhangrunwu@bit.edu.cn}
\affiliation
{Centre for Quantum Physics, Key Laboratory of Advanced Optoelectronic Quantum Architecture and Measurement (MOE), School of Physics, Beijing Institute of Technology, Beijing, 100081, China}

\author{Botao Fu}
\email[]{fubotao2008@gmail.com}
\affiliation{College of Physics and Electronic Engineering, Center for Computational Sciences, Sichuan Normal University, Chengdu, 610068,
P.R.China}
\date{\today}

\begin{abstract}
The interplay between magnetic ordering and band topology has emerged as a fertile ground for discovering novel quantum states with profound implications for fundamental physics and next-generation electronics. Here, we theoretically predict a new type-III Weyl semi-half-metal (SHM) state in monolayer Li$_2$N, uniquely combining magnetic half-metallicity and type-III Weyl semimetal characteristics. First-principles calculations reveal a fully spin-polarized and critically tilted Weyl cone around the Fermi level in monolayer Li$_2$N, driven by $p$-orbital ferromagnetism. This arises from the symmetry-protected band crossing between a flat valence band and a highly dispersive conduction band, leading to type-III Weyl fermions with strong transport anisotropy. A low-energy $k{\cdot}p$ Hamiltonian is constructed and corresponding nontrivial edge states are uncovered to capture the topological nature of Li$_2$N.
Notably, this Weyl SHM phase remains robust under biaxial strain ranging from -2$\%$ to $4\%$, with an ideal type-III Weyl fermion emerging alongside a line-like ergodic surface emerging at 3.7$\%$ strain, offering a promising platform for exploring correlated electronic phenomena.
Our results establish Li$_2$N as a viable candidate for realizing exotic type-III Weyl SHM states and open a new avenue for exploring the intricate interplay among magnetism, topology, and flat-band physics.
\end{abstract}

\maketitle

\section{ introduction}
Recent advances in topological magnetic materials have opened new avenues for exploring novel quantum states with promising applications in topological spintronic applications\cite{zhang2023magnetic,zou2019study,xu2020high,PhysRevX.12.021016}. In particular, the combination of ferromagnetic half-metallicity with topological semimetallicity has given rise to a new class of quantum materials, topological semi-half-metals (SHMs)\cite{wang2020spin,zhang2021weyl,PhysRevB.99.075131}.
These materials inherit the fully spin polarization characteristic of half-metals while simultaneously exhibiting the nontrivial topological features of semimetals, making them highly promising for spintronic and quantum transport applications\cite{PhysRevLett.124.016402,PhysRevLett.121.246401,PhysRevLett.129.036801}.

Topological SHMs can be systematically classified based on their band structures and nodal features\cite{RevModPhys.90.015001,RevModPhys.93.025002}.
Following the classification of nonmagnetic topological semimetals, SHMs are categorized into 0D Weyl/Dirac SHMs \cite{PhysRevLett.123.206601,PhysRevB.109.174426,PhysRevMaterials.3.021201}, 1D nodal-line SHMs \cite{PhysRevB.99.035125}, and potentially higher-dimensional nodal-surface SHMs\cite{PhysRevB.103.195115}.
Additionally, they can be distinguished by the tilt of their Dirac or Weyl cones into type-I, over-tilted type-II, and critically tilted type-III cases.
Type-I Weyl Semimetals (WSMs) feature point-like Fermi surfaces\cite{PhysRevB.83.205101}, while type-II WSMs\cite{soluyanov2015type} possess hyperbolic Fermi surfaces (touched electron and hole pockets), leading to exotic transport properties such as large magnetoresistance and unconventional optical responses \cite{PhysRevB.95.094513,PhysRevB.106.165404,PhysRevB.105.205203}.
More intriguingly, type-III WSMs represent the ``critical transition state" between type-I and type-II, characterized by line-like Fermi surface and flat bands that have drawn significant theoretical attention\cite{PhysRevB.110.125204,tan2024bipolarized,PhysRevB.101.100303,PhysRevB.103.L081402,PhysRevMaterials.7.014202}. For instance, their quasiparticles can simulate phenomena such as black-hole horizon in quantum gravity in quantum gravity theories\cite{PhysRevB.98.121110,sims2023analogous}.

Despite significant theoretical advancements, experimental realization of topological SHMs, particularly type-III systems, remains challenging.
While type-I and type-II SHMs have been predicted in various magnetic materials\cite{PhysRevB.100.064408,wei2016spin,gong2024genuine,PhysRevB.100.064408, zhou2019fully}, the realization of type-III SHMs have been largely unexplored. To date, type-III fermions has been primarily proposed in photonic\cite{PhysRevX.9.031010} and limited electronic systems\cite{fragkos2021type,kang2020topological}, with no examples identified in ferromagnetic half-metals, likely due to the stringent material and symmetry requirements needed to stabilize such exotic states\cite{mizoguchi2020type}.

Motivated by this gap in this field, we propose the concept of type-III Weyl SHM by combining the intrinsic spin polarization of half-metals with the critically tilted type-III Weyl fermions. Using first-principles calculations, we predict that monolayer Li$_2$N hosts such type-III Weyl fermions within a single spin channel. This quasi-fermionic state arises from the non-trivial band crossing between a nearly flat valence band and a highly dispersive conduction band, primarily derived from the N's $p_z$ and $p_{x,y}$ orbitals, respectively. The extreme anisotropy of type-III fermion results in a highly tunable Fermi velocity, ranging from nearly zero to 7.48$\times 10^5$ m/s. Moreover, biaxial strain not only effectively modulates the position and energy of the Weyl fermion but also plays a crucial role in tuning the band dispersion. Remarkably, at a critical tensile strain of 3.7$\%$, a perfect type-III Weyl fermion state emerges, characterized by an exact flat band and a line-like ergodic surface.
Our findings provide the first material candidate for type-III Weyl SHM and establish a novel platform for investigating the interplay between magnetism, topology, and flat-band physics.

\section{Computational Methods}

The first-principles calculations were conducted using the Vienna \textit{ab initio} Simulation Package (VASP) \cite{PhysRev.136.B864,PhysRev.140.A1133,kresse1996efficiency}, based on density functional theory (DFT). The exchange-correlation functional was treated within the generalized gradient approximation (GGA) using the Perdew-Burke-Ernzerhof (PBE) functional \cite{PhysRevB.54.11169,PhysRevLett.77.3865}. The interaction between electrons and ions was described using the projector augmented-wave (PAW) method.
For the Li$_2$N monolayer, a vacuum space of 15~{\AA} was introduced to eliminate potential interactions between periodic layers. A cutoff energy of 600~eV was employed for the plane-wave basis set. The Brillouin zone was sampled using a Monkhorst-Pack $k$-mesh with a size of $15 \times 15 \times 1$. The convergence criteria for force and energy were set to 0.001~eV/{\AA} and $10^{-7}$~eV, respectively. To evaluate the dynamical stability of the Li$_2$N monolayer, the phonon spectrum was calculated using the phonopy package \cite{PhysRevB.78.134106}. Additionally, the topological edge states were using the WANNIERTOOLS package\cite{wu2018wanniertools}.

\section{Structure and stability of monolayer Li$_2$N}
\begin{figure}
	\includegraphics[width=9.0cm]{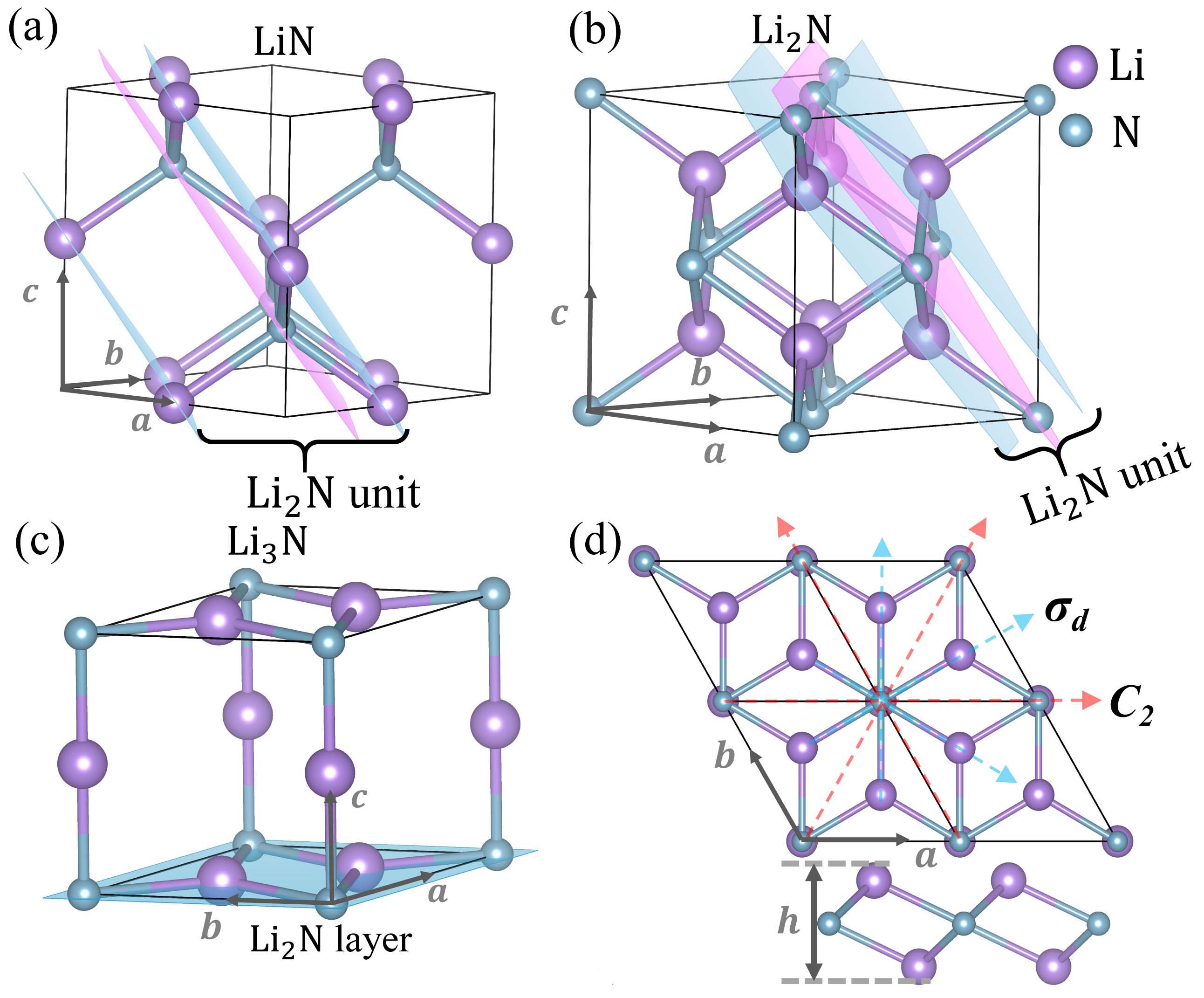}
	\centering
 \caption{(a) The Bulk LiN with SG. of F$\overline{4}$3m. (b) The bulk Li$_2$N with SG. of Fm$\overline{3}$m), (c) The bulk Li$_3$N with SG. of P6/mm, (d) The monolayer Li$_2$N with SG. of P$\overline{3}$m1, which can be exfoliated from bulk Li$_3$N, Li$_2$N, and LiN, respectively.}
    \label{Fig1}
\end{figure}

Three-dimensional (3D) lithium nitrides, including Li$_x$N ($x$ = 1, 2, 3) \cite{wang2012lithium,laniel2018direct,ostlin2016density,tapia2020low}, have been experimentally synthesized. As shown in Figs.~\ref{Fig1}(a)-(c), bulk LiN \cite{laniel2018direct} and Li$_2$N \cite{wang2012lithium} form face-centered cubic lattices and exhibit magnetic metallic behavior, while the bulk Li$_3$N \cite{ostlin2016density} belongs to a non-magnetic semiconductor with a hexagonal structure.
The monolayer Li$_2$N, depicted in Fig.~\ref{Fig1}(d), has a 1T-MoS$_2$-like configuration with space group (SG.) of $P\overline{3}m1$ and point group $D_{3d}$. It features a triangular lattice of N atoms sandwiched between two layers of Li atoms, with optimized lattice constants $\textit{\textbf{a}}$=$\textit{\textbf{b}}$=3.153~{\AA}, Li-N bond length $l$=2.010~{\AA}, and layer height $h$=1.702~{\AA}. Each N atom is trigonally coordinated to six Li atoms, forming a trigonal prism. This structure breaks horizontal mirror symmetry ($M_z$) but hosts spatial inversion ($P$), three two-fold rotation axes ($C_2$) and reflection planes ($\sigma_d$) related by $C_{3z}$.
The stability of monolayer Li$_2$N is confirmed by the calculations of phonon dispersion, $ab$ $initio$ molecular electrodynamics as well as mechanical constants\cite{PhysRevB.105.075402} as shown in Fig.~S1 of the supporting information (SI).

In addition, our calculations indicate that such monolayer Li$_2$N can be obtained by breaking the interlayer ionic bonds along the (111) direction of bulk Li$_2$N in Fig.~\ref{Fig1}(b). Similarly, as shown in Fig.~\ref{Fig1}(c), exfoliating bulk Li$_3$N along the (001) direction can isolate a planar Li$_2$N layer, which further spontaneously reconstructs into the 1T-Li$_2$N structure upon relaxation. Likewise in Fig.~\ref{Fig1}(a), breaking the interlayer ionic bonds of bulk LiN along the (111) direction yields a Li$_2$N unit that also spontaneously reconstructs into the  1T-Li$_2$N phase.
These findings indicate that, as a 2D limitation of all Li$_x$N, 1T-Li$_2$N is energetically favorable. Notably, similar non-layered two-dimensional (2D) materials, such as WO$_3$ \cite{tao2019recent} and NaCl$_2$ \cite{yi2025two}, have been successfully synthesized using techniques like liquid-metal-assisted exfoliation and cryogenic synthesis. Therefore, we propose that monolayer 1T-Li$_2$N can be synthesized through exfoliation from its experimentally available non-layered Li$_x$N precursors.

\section{Half-metallicity of monolayer Li$_2$N}
A striking feature of monolayer Li$_2$N is its unconventional stoichiometry, where each Li atom donates one electron, and the N atom, acquiring two electrons, adopts a -2 oxidation state, as shown in Fig.~S2(a) of SI.
According to Hund's rules, this electron configuration results in an uncompensated spin, leading to local magnetic moments in monolayer Li$_2$N. This phenomenon is confirmed through our DFT calculations by systematically evaluating all possible magnetic configurations, including non-magnetic (NM), ferromagnetic (FM), and antiferromagnetic (AFM) states, as presented in the Fig.~S3 of SI. By comparing their total energies, we identify the FM state as the ground state, with a total magnetic moment of 1.0 $\mu_B$ per unit cell. Notably, the FM state is 0.265 eV lower in energy than the AFM state and 0.087 eV lower than the NM state.

\begin{figure}
	\includegraphics[width=9cm]{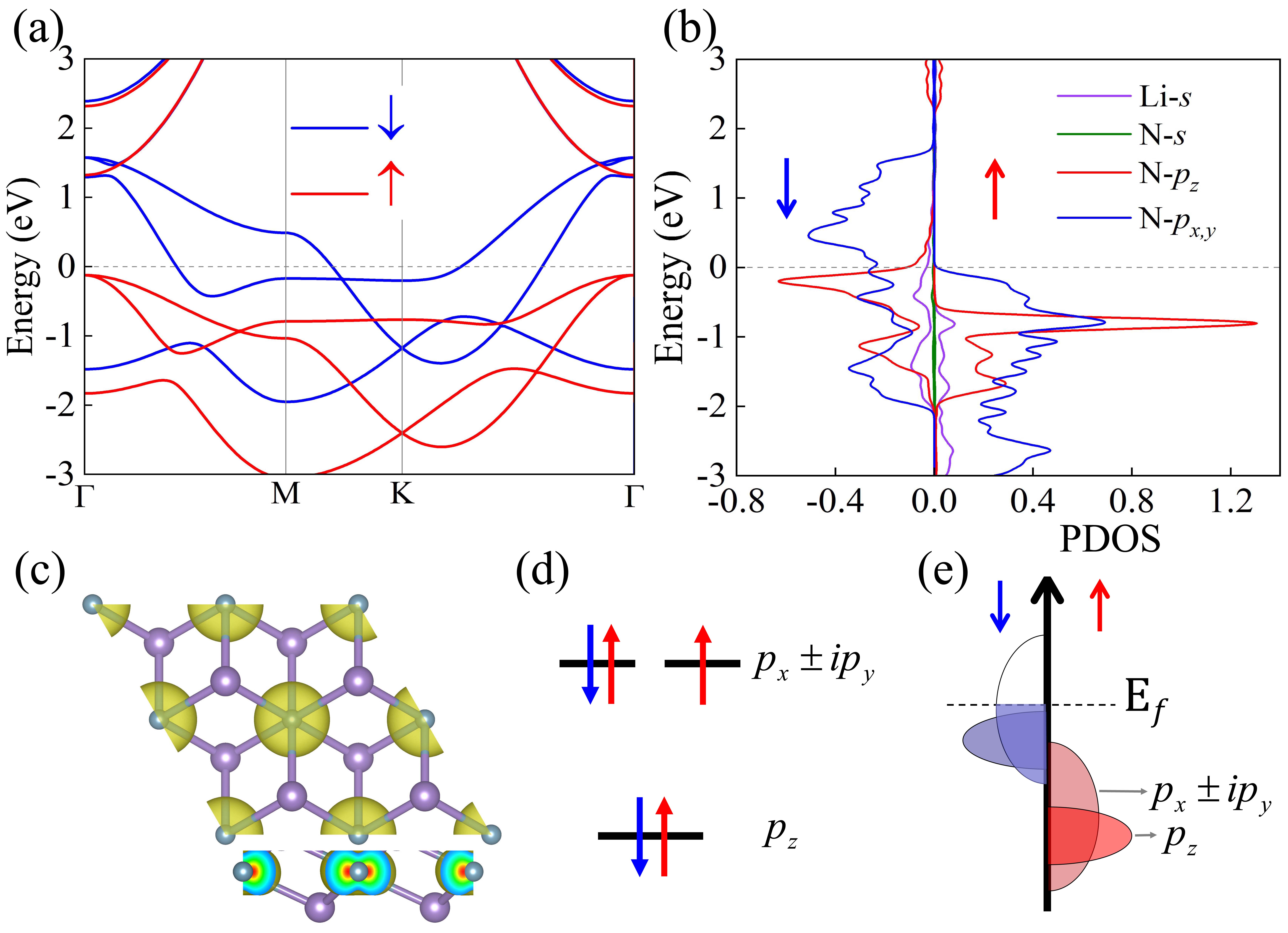}
	\centering
    \caption{ (a) The spin-polarized band structure of monolayer ${\rm{Li_2N}}$, with red and blue bands representing the spin-up and spin-down channels, respectively. (b) The spin-resolved PDOS projected on different orbitals of monolayer ${\rm{Li_2N}}$.
    (c) The spin charge charge density distribution of monolayer ${\rm{Li_2N}}$. (d) Schematic illustration of $p$-orbital splitting and electron filling in monolayer ${\rm{Li_2N}}$. (e) Schematic representation of band broadening and exchange-field-induced spin splitting in monolayer ${\rm{Li_2N}}$.}
    \label{Fig2}
\end{figure}

The spin-polarized band structure of FM Li$_2$N is shown in Fig.~\ref{Fig2}(a). The spin-up (minority) channel features a sizable half-metallic gap of 1.45~eV, while the spin-down (majority) channel remains metallic, ensuring 100\% spin polarization. This characteristic firmly classifies Li$_2$N as a half-metal. The spin-density distribution, illustrated in Fig.~\ref{Fig2}(c), reveals that the magnetic moment is predominantly localized on the N atoms, with dominant contributions from the $p_x{\pm}ip_y$ orbitals.

This behavior can be understood in terms of crystal-field splitting within the $D_{3d}$ symmetry. As depicted in Fig.~\ref{Fig2}(d), the three degenerate $p$-orbitals of the N atom split into a nondegenerate $p_z$ orbital and two degenerate $p_x \pm ip_y$ orbitals under a $D_{3d}$ crystal field \cite{wang2016quantum,hua2022topological}. In monolayer Li$_2$N, each N atom receives two electrons from adjacent Li atoms, forming N$^{2-}$ ions with seven outer-shell electrons. Among these, two electrons occupy the $s$-orbital, two reside in the $p_z$-orbital, while the remaining three fill the degenerate $p_x \pm ip_y$ orbitals, resulting in a net magnetic moment of 1.0 $\mu_B$.

Within the periodic crystal lattice of Li$_2$N, the discrete $p$-orbital energy levels broaden into energy bands. When an exchange field is introduced, it induces spin splitting of these bands, as depicted in Fig.~\ref{Fig2}(e). If this exchange splitting exceeds the band broadening, the system attains a half-metallic state.
This conclusion is further substantiated by the partial density of states (PDOS) in Fig.~\ref{Fig2}(b), which shows that the N's $p_z$ orbital is fully occupied in both the spin-up and spin-down channels, thus making no contribution to the net magnetization. In contrast, the $p_x \pm ip_y$ orbitals are fully occupied in the spin-up channel but only partially filled in the spin-down channel. Consequently, the half-metallic ferromagnetism in Li$_2$N primarily originates from the partially occupied $p_x \pm ip_y$ orbitals of the N atom, in agreement with the results in Fig.~\ref{Fig2}(b).

\section{Emergent type-III Weyl Semi-half-metal in monolayer Li$_2$N}
\begin{figure*}
	\includegraphics[width=17cm]{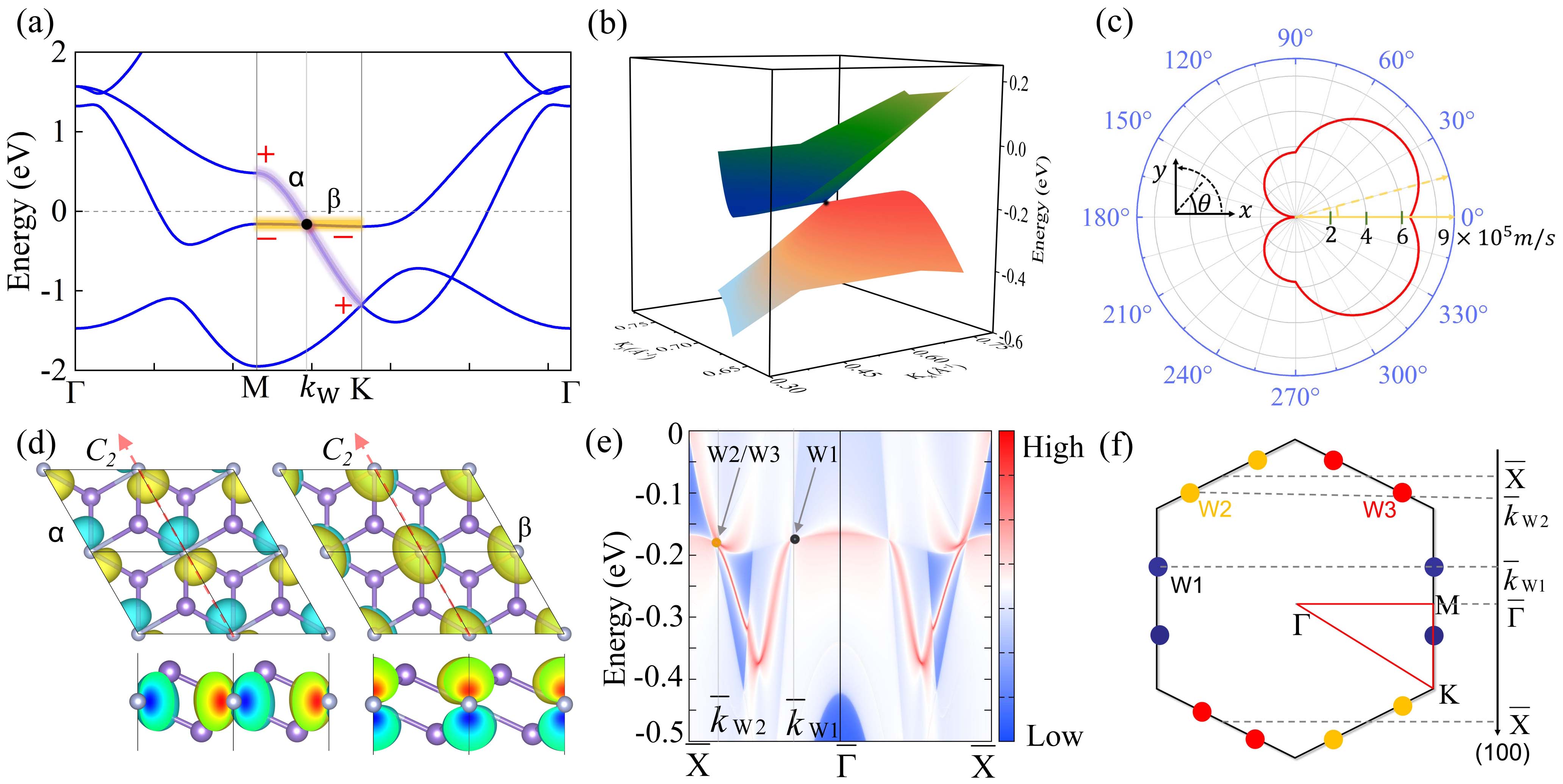}
	\centering
    \caption{ (a) Enlarged of spin down band structure near the Fermi level. The conduction and valence bands along MK path are highlighted and labeled as $\alpha$ and $\beta$. The signs above indicate the eigenvalues of Bloch functions for the $C_2$ rotation operation.
(b) 3D band structure around the WP.
(c) Angular-dependent Fermi velocity around the WP.
(d) The Bloch functions of bands $\alpha$ and $\beta$.
(e) Edge states for the semi-infinite nanoribbon of monolayer Li$_2$N along $x$ direction.
(f) Distribution of three pairs of WPs (W1, W2 and W3) in the Brillouin zone, along with the one-dimensional projected Brillouin zone along the $x$ direction.}
    \label{Fig3}
\end{figure*}

Here, we focus on the topological properties of monolayer Li$_2$N, as shown in Fig.~\ref{Fig3}(a).
Firstly, near the Fermi level, two fully spin-polarized bands, labeled as $\alpha$ and $\beta$, cross each other along the MK path, forming spin-polarized Weyl point (WP). Further analysis of the Bloch function symmetries reveals that $\alpha$ and $\beta$ bands possess opposite eigenvalues of $C_2$ rotation, indicating that these WPs are protected by the two-fold rotation symmetry $C_2$. Analogous to 3D Dirac semimetals protected by crystalline symmetry\cite{PhysRevB.88.125427}, the emergence of such symmetry-protected 2D WPs can also be understood through band inversion analysis. We define a local band gap as:
\begin{eqnarray}
\Delta E(\textit{\textbf{k}}) = E_{\alpha}(\textit{\textbf{k}}) - E_{\beta}(\textit{\textbf{k}}). \label{dEk}
\end{eqnarray}
Along the MK path, it is evident that $\Delta E(\mathrm{M})>0$ and $\Delta E(\mathrm{K})<0$ . This indicates that a band inversion occurs along this path, which implies the existence of a band crossing point, satisfying the condition $\Delta E(\textit{\textbf{k}}) = 0$ at a certain $\textit{\textbf{k}}$ point.
Furthermore, considering the overall $D_{3v}$ symmetry of the system, there exist three pairs of such spin-polarized WPs (labelled as W1, W2, W3) as shown in Fig.~\ref{Fig3}(f).

Secondly, the $\alpha$ and $\beta$ bands exhibit strikingly different dispersion characteristics along the MK path. Band-$\alpha$ is highly dispersive, featuring a broad bandwidth of 1.64 eV and an exceptionally high Fermi velocity on the order of $10^5$~m/s. In contrast, band-$\beta$ exhibits a prototypical flat-band nature, with an extremely narrow bandwidth of just 0.02 eV.
This unusual crossing between a highly dispersive band and an almost flat band gives rise to the rare type-III Weyl SHM state in Li$_2$N.

To further elucidate the origin of this sharp contrast, we analyze the corresponding Bloch wave functions of band-$\alpha$ and band-$\beta$, as shown in Fig.\ref{Fig3}(d). Therein, band-$\alpha$ exhibits in-plane $p$-orbital characteristics, with wave functions extending along the direction of $C_2$ axis, whereas band-$\beta$ is primarily composed of localized $p_z$ orbital. These distinct orbital characteristics fundamentally determine their drastically different dispersion behaviors along the MK direction.
Additionally, their wave functions exhibit opposite symmetries under the $C_2$ operation, which is fully consistent with the direct eigenvalue calculations presented in Fig.\ref{Fig3}(a).

To better capture the underlying physics of type-III Weyl SHM in Li$_2$N, a spin-polarized low-energy effective $k{\cdot}p$ Hamiltonian is developed as\cite{fu20232d},
\begin{eqnarray}
H(\textit{\textbf{q}}) =  v_x q_x \sigma_x + v_y q_y \sigma_y +w_{x} q_x \sigma_0,\label{kpmodel}
\end{eqnarray}
where $\textit{\textbf{q}}$ =($q_x$, $q_y$) denote the wave vector relative to the WP, $\sigma_0$ is the identity matrix, and $\sigma_{x,y}$ are Pauli matrices. The parameters $v_x=3.183 \times 10 ^5 m/s$, $v_y=3.677 \times 10 ^5 m/s$, and $w_x=3.304 \times 10 ^5 m/s$ are obtained by fitting $k{\cdot}p$ bands with DFT results. The eigenenergies of this Hamiltonian are given by,	
\begin{eqnarray}
E_{\pm}(\textit{\textbf{q}}) = w_{x} q_x \pm \sqrt{(v_x q_x)^2 + (v_y q_y)^2}. \label{kpband}
\end{eqnarray}
This describes a Weyl cone tilted along $k_x $-direction, with the degree of tilting characterized by the ratio $r_t = {w_x}/{v_x}$.
According to classification criteria, type-I Weyl fermions correspond to $r_t < 1$, while type-II Weyl fermions correspond to $r_t > 1$. The type-III Weyl fermion represents the critical transition state between type-I and type-II, defined by $r_t = 1$.

For monolayer Li$_2$N, the calculated value $r_t = 1.038$ is very close to unity, indicating a quasi-type-III Weyl state with highly anisotropic Fermi velocity. As shown in Fig.~\ref{Fig3}(c), the Fermi velocity reaches $7.48 \times 10^5$ m/s along ${\theta} \approx 30^\circ$, surpassing most known 2D magnetic Weyl semimetals \cite{wang2013strain, zhang2015spin, cai2015single}, such as 1T-YN$_2$ ($3.7 \times 10^5$ m/s) \cite{liu2017yn} and VCl$_3$ ($1.6 \times 10^5$ m/s) \cite{he2016unusual}. In contrast, along the $-k_x$ direction, the Fermi velocity is drastically reduced to just $0.22 \times 10^5$ m/s.
These findings firmly establish Li$_2$N as a rare realization of the type-III Weyl SHM phase.

As is well known, for 2D Weyl SHM, topological edge states connecting a pair of WPs exist in a single spin channel, as dictated by the bulk-boundary correspondence. Based on the Wannier-based tight-binding (TB) model\cite{wu2018wanniertools,mostofi2008wannier90}, we have calculated the electronic structure of a semi-infinite nanoribbon of Li$_2$N. As shown in Fig.~\ref{Fig3}(f), along the (111) direction, one pair of WPs (W1) projects onto the $\overline{k_{W1}}$ point, while the other two pairs (W2 and W3) project onto the same $\overline{k_{W2}}$  point. In Fig.~\ref{Fig3}(e), non-trivial topological edge states connecting $\overline{k_{W1}}$ and $\overline{k_{W2}}$  can be observed, thereby confirming the topological nature of the Li$_2$N system.

\section{Strain-induced topological phase transition in Li$_2$N}
\begin{figure*}
	\includegraphics[width=17.5 cm]{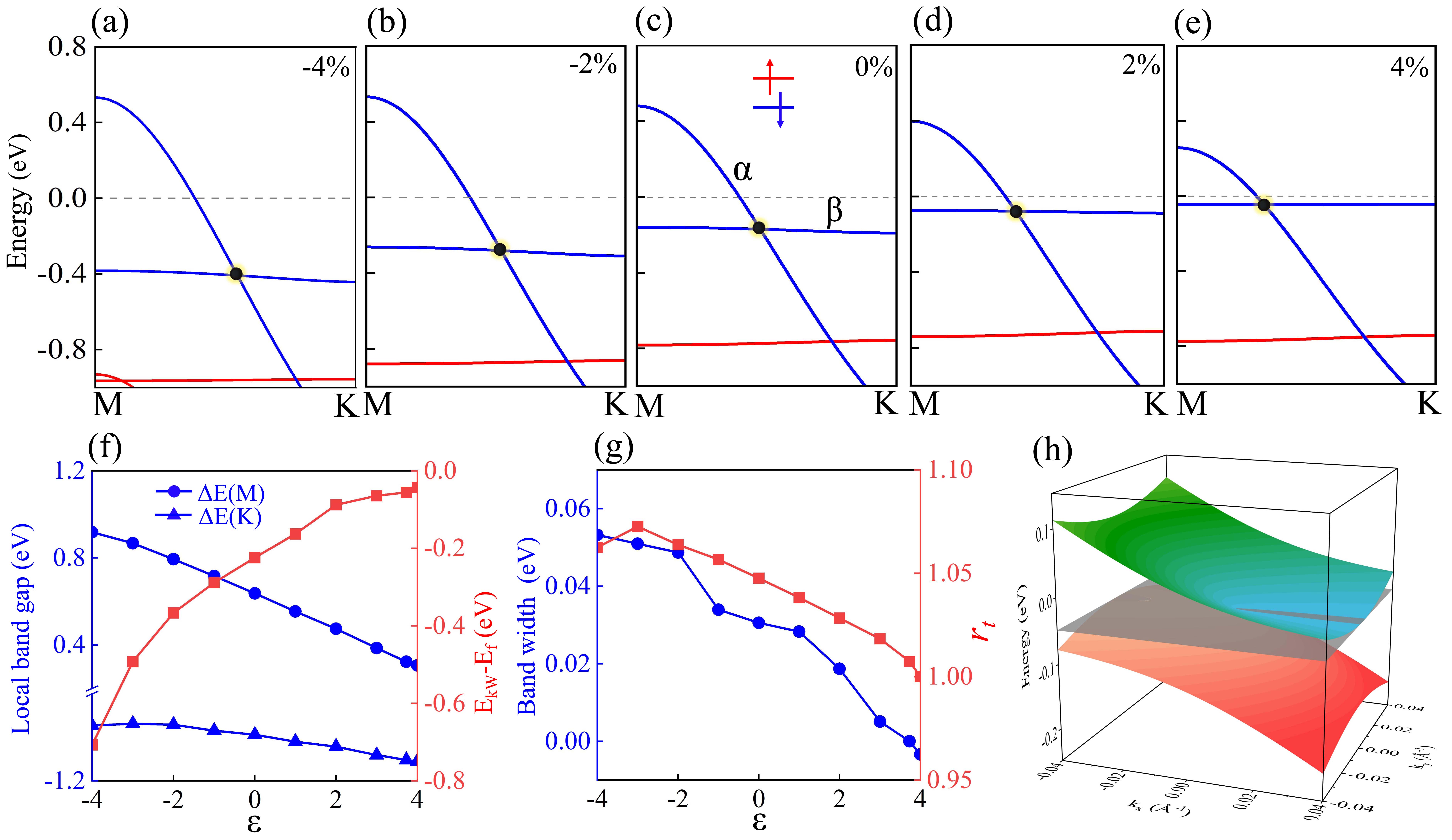}
	\centering
    \caption{(a)-(e) The spin-polarized band structures of monolayer Li$_2$N under biaxial strains, ranging from $-4\%$ to $4\%$.
(f) The local band gap of M and K point, the energy of Weyl fermion respect to Fermi level as a function of strain.
(g) The band width of band-$\beta$ and the tilting degree ($r_t$) of Weyl fermion as a function of strain.
(h) 3D view of ideal type-III Weyl fermion under 3.7$\%$ biaxial strain.}
    \label{Fig4}
\end{figure*}

Strain plays a crucial role in modulating atomic interactions, significantly influencing a material¡¯s magnetic, electronic, and topological properties\cite{PhysRevB.102.125118,PhysRevB.108.224417,liu2025study,babalola2024investigation}. Here, we explore the strain effect on the Weyl SHM state in monolayer Li$_2$N. Biaxial strain is defined as $\varepsilon = ({a - a_0})/{a_0}$, where $a$ and $a_0$ are the lattice constants of the strained and unstrained structures, respectively. A negative $\varepsilon$ indicates a compressive state, while a positive $\varepsilon$ indicates a tensile state. A biaxial strain ranging from -4$\%$ to 4$\%$ is considered. The electronic structure and magnetic properties of the system were computed, as shown in Fig.~S4. Our results reveal that both spin polarization energy and half-metallic gap increase with the lattice constant, indicating that tensile strain can enhance the ferromagnetic half-metallic nature (Fig.~S4(f) in SI). However, when compressive strain exceeds 2$\%$, the spin-up channel also crosses the Fermi level, reducing the total magnetic moment to below 1.0 $\mu_B$ and disrupting the half-metallic character (Fig.~S4(g) in SI). Thus, monolayer Li$_2$N retains its ferromagnetic semimetallic state within the strain range of -2$\%$ to 4$\%$.

We next examine the strain-dependent evolution of the band structure in monolayer Li$_2$N. As compressive strain decreases or tensile strain increases, both $\Delta E(\mathrm{M})$ and $\Delta E(\mathrm{K})$ exhibit a decreasing trend as shown in Fig.~\ref{Fig4}(f). However, their opposite signs remain unchanged throughout the process, indicating the robustness of the WP along the MK path against external strain.
More intriguingly, as the $\Delta E(\mathrm{M})$ gradually decreases, the position of WP shift toward the M point, and its energy moves closer to the Fermi level, making them highly favorable for experimental observation.

Furthermore, the flat-band characteristic of band-$\beta$ and the the highly dispersive nature of band-$\alpha$ in Li$_2$N remain remarkably stable under external strain, perserving the type-III Weyl fermions. As illustrated in Fig.~\ref{Fig4}(g), the tilt parameter $r_t$ exhibits only minor variations around 1.0, ranging from 1.07 to 0.99, which confirms the stability of type-III nature of Weyl fermion.
Notably, under increasing tensile strain, band-$\beta$ becomes progressively flatter. At a critical strain of 3.7$\%$, $r_t$ reaches exactly 1.0, giving rise to a perfectly flat band (Fig.~S4(h) in SI) along with a line-like ergodic surface state, as shown in Fig.~\ref{Fig4}(h). This transition signifies the realization of an ideal type-III Weyl SHM.
Thus, the tunability of the half-metallic phase, combined with the exceptional stability of the flat-band feature over a broad strain range, ensures the stable existence of fully spin polarized type-III Weyl fermions in Li$_2$N, further enhancing their potential for experimental realization.

\section{Discussion And Conclusion}
To ensure the accuracy of our findings, we further refined the band structure using the HSE06 functional\cite{krukau2006influence}. Our results confirm the persistence of two distinct bands, $\alpha$ and $\beta$, near the Fermi level, which intersect along the MK path, thereby preserving the unique features of type-III Weyl HSM, as shown in Fig.~S2(b) of SI.
Additionally, by calculating the magnetic anisotropy energy (MAE) in both the in-plane and out-of-plane directions, we found that the easy magnetization axis lies within the plane, with the out-of-plane direction being 10 $\mu$eV higher in energy, as shown in Fig.~S5(a). Within the plane, the system exhibits nearly isotropic behavior, with energy differences between different in-plane directions not exceeding 0.5 $\mu$eV. The lowest-energy magnetization direction is identified as the (100) direction, as shown in Fig.~S5(b).
Although Li and N are both light elements, we also examined the impact of spin-orbit coupling (SOC) on the system's topological properties. With SOC included and the magnetization aligned along the (100) direction, two pairs of WPs (W1/W3) develop a small band gap of 5 meV, while the remaining pair of WPs (W2) along the (100) direction remains protected by the two-fold rotational symmetry, as demonstrated in Figs.~S5(c)-(e). This confirms that even with SOC effects, the system retains its type-III WHSM nature.

In summary, our study has unveiled the unique electronic and topological properties of monolayer Li$_2$N, establishing it as a novel 2D type-III Weyl SHM with significant potential for spintronic applications.
The interplay between flat valence bands and highly dispersive conduction bands leads to exotic Weyl fermions that differ fundamentally from those reported in previous materials. This discovery not only expands the family of 2D topological materials but also provides critical insights for designing future electronic and quantum devices. Further experimental validation and exploration of its functional applications could pave the way for innovative advancements in spintronics and topological electronics

\begin{acknowledgements}
This work was supported by the Natural Science Foundation of China (Grants No. 12204330 and No. 12304188), the Natural Science Foundation of Beijing (Grant No. 1252029), the Beijing National Laboratory for Condensed Matter Physics (Grant No. 2024BNLCMPKF006). B. Fu acknowledges financial support from Sichuan Normal University (Grant No. 341829001). The numerical computations were conducted at the Hefei Advanced Computing Center, with additional support from the High-Performance Computing Center of Sichuan Normal University, China.

\end{acknowledgements}

\end{document}